\documentclass{Interspeech}

\newcommand{\mmedit}[1]{\textcolor{black}{#1}} %
\usepackage{multirow}
\usepackage{amsmath,graphicx}
\usepackage{url}

\interspeechcameraready

\title{Speaker-agnostic Emotion Vector for Cross-speaker Emotion Intensity Control}

\author[affiliation=1]{Masato}{Murata}
\author[affiliation=1]{Koichi}{Miyazaki}
\author[affiliation=1]{Tomoki}{Koriyama}

\affiliation{}{CyberAgent}{Japan}
\email{murata\_masato@cyberagent.co.jp, miyazaki\_koichi\_xa@cyberagent.co.jp, koriyama\_tomoki@cyberagent.co.jp}

\usepackage{comment}

\begin{document}

\maketitle
\begin{abstract}
Cross-speaker emotion intensity control aims to generate emotional speech of a target speaker with desired emotion intensities using only their neutral speech. A recently proposed method, emotion arithmetic, achieves emotion intensity control using a single-speaker emotion vector. 
Although this prior method has shown promising results in the same-speaker setting, it lost speaker consistency in the cross-speaker setting due to mismatches between the emotion vector of the source and target speakers.
To overcome this limitation, we propose a speaker-agnostic emotion vector designed to capture shared emotional expressions across multiple speakers.
This speaker-agnostic emotion vector is applicable to arbitrary speakers.
Experimental results demonstrate that the proposed method succeeds in cross-speaker emotion intensity control while maintaining speaker consistency, speech quality, and controllability, even in the unseen speaker case.

\end{abstract}

\section{Introduction}
Emotion intensity control aims to generate emotional speech with desired emotion intensity, such as ``strong angry'' or ``weak sad.''
This capability is useful for applications like conversational assistants, audio advertisements, and audiobook narrations.
Although several methods have been proposed~\cite{mixed_emotion, FET, iemotts, lei2022msemotts}, they typically require substantial amounts of emotional speech data of the target speaker.
\mmedit{However, obtaining specific emotional speech data of the target speaker is not always easy in practical usage.}
To address this issue, several studies have focused on the task called ``cross-speaker emotion intensity control,'' which enables emotion intensity control using only the target speaker’s neutral (reading-style) speech.
\mmedit{A key in this task is disentangling emotion and speaker representations. }

\begin{figure}[t]
  \centering
  \vspace{15pt} %
   \begin{minipage}[b]{1.0\linewidth}
     \centering
     \includegraphics[width=\hsize]{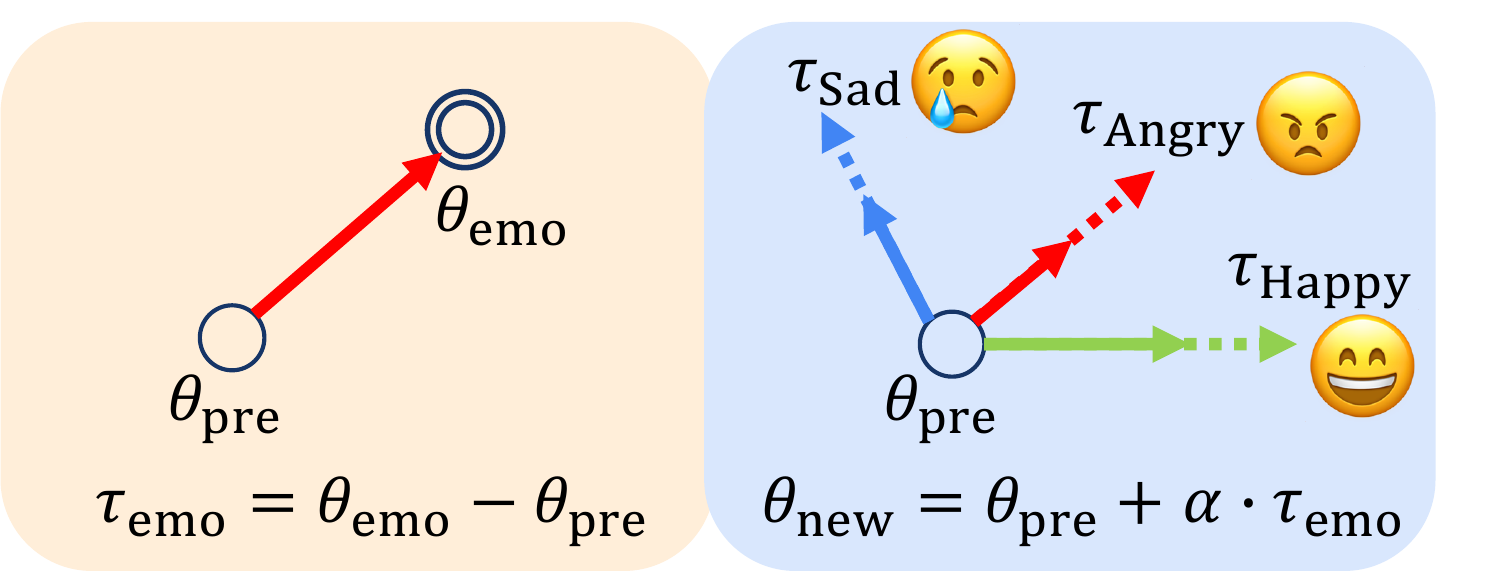}
   \end{minipage}
   \vspace{-15pt} %
   \caption{Emotion intensity control through emotion vector.}
   \vspace{-15pt} %
   \label{fig:overview}
 \end{figure}

\looseness=-1
\mmedit{For cross-speaker emotion intensity control, }
Li et al.~\cite{crossspeaker} proposed an emotion disentangling module that separates the speaker identity from emotion embeddings using specifically designed loss functions for cross-speaker emotion transfer.
Similarly, Jo et al. introduced tailored loss functions 
to disentangle emotion and speaker information~\cite{crossspeaker2}.
EmoSphere++ also achieved cross-speaker emotion intensity control task by employing an emotion-adaptive spherical vector to represent continuous emotion variations, combined with a multi-level style encoder~\cite{emospherepp}.
In contrast, 
Kalyan et al. proposed emotion arithmetic~\cite{emotion_arithmetic}, which achieved emotion intensity control by directly editing the trained model parameters without relying on any additional modules or loss functions.
This method is based on task arithmetic~\cite{task_arithmetic}, which manipulates model behaviors using a ``task vector'' derived by subtracting pre-trained model parameters from fine-tuned ones.
\mmedit{Specifically, emotion arithmetic builds an ``emotion vector'' by subtracting the model parameters of a neutral text-to-speech (TTS) model (pre-trained) from those of an emotional TTS model (finetuned).
This emotion vector is expected to capture the specific emotion expression information.}
To control emotion intensity, the scaled emotion vector is added to the pre-trained model depending on the scaling factor, as shown in Figure~\ref{fig:overview}.
Moreover, they showed that an emotion vector obtained from one speaker could be applied to other speakers' TTS models, enabling cross-speaker emotion intensity control.
However, experimental results reported that in this cross-speaker setting, the generated speech lost speaker consistency.
Since these emotion vectors are derived from a single speaker, they reflect the speaker-specific emotional expression. Consequently, the discrepancies between the emotion vectors of the source and target speakers cause the failure in generating cross-speaker emotion speech.

\looseness=-1
To overcome this limitation, we propose a speaker-agnostic emotion vector that improves speaker consistency and speech quality in cross-speaker emotion intensity control.
This speaker-agnostic emotion vector is designed to capture the shared emotional expressions across multiple speakers.
Therefore, this method can be applied to arbitrary speakers for cross-speaker emotion intensity control while maintaining speaker consistency.
Moreover, several previous studies~\cite{yourtts, xvector, scglow_tts} have shown that using x-vector speaker conditioning
enables models to generate an unseen speaker's speech. 
Based on this, we adopted x-vector conditioning in our proposed method to achieve emotion intensity control even for unseen speakers.
We evaluated the proposed method in terms of speech quality, speaker consistency, and emotion intensity controllability.
Experimental results demonstrate that the proposed method succeeds in cross-speaker emotion intensity control while maintaining speaker consistency, speech quality, and controllability, even in the unseen speaker case.

Audio samples are available on our demo page\footnote{\url{https://muramasa2.github.io/speaker-agnostic-emotion-vector/}}.

\begin{figure}[t]
 \centering
  \begin{minipage}[b]{1\linewidth}
    \centering
    \includegraphics[width=\hsize]{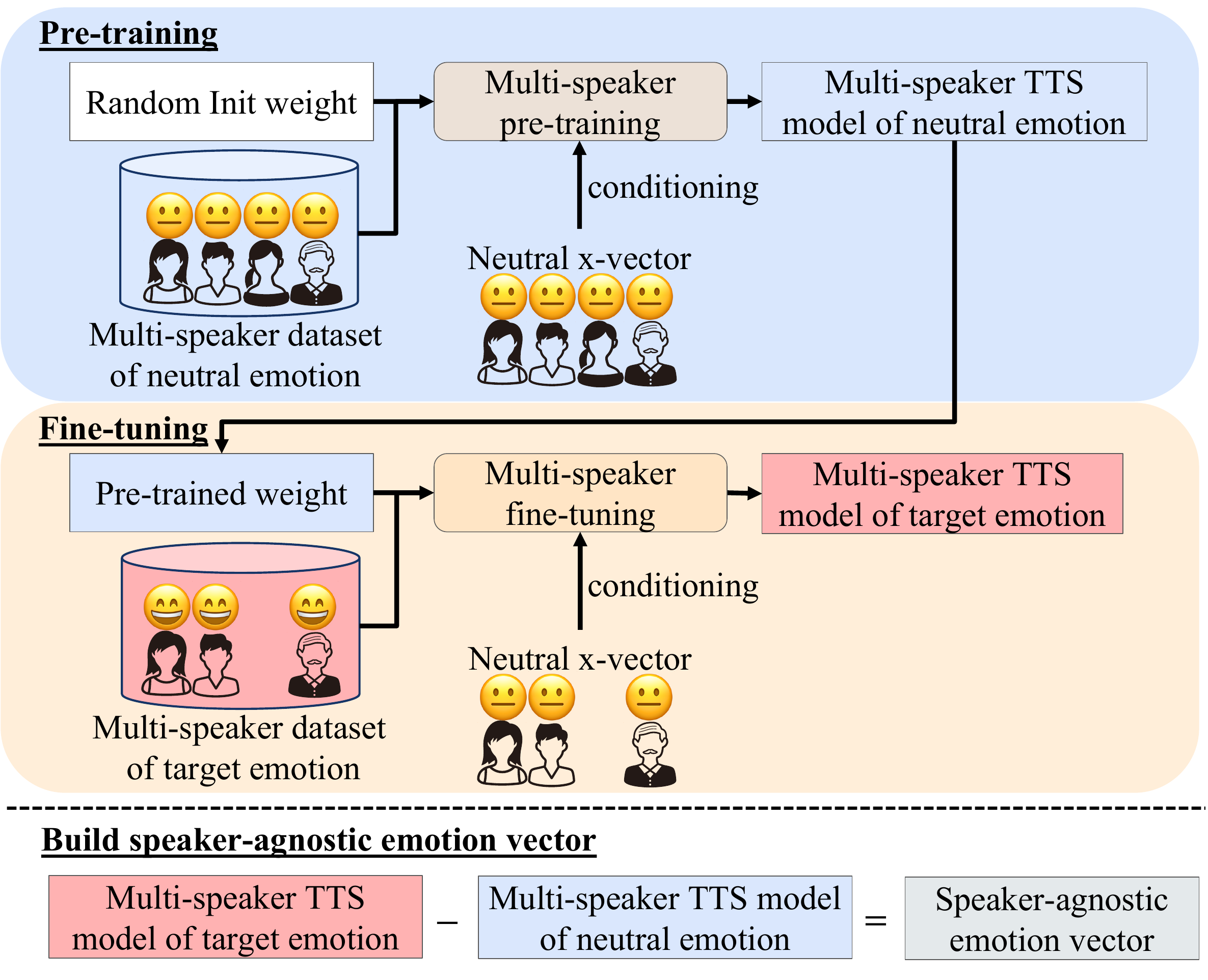}
  \end{minipage}
  \caption{The process of creating the speaker-agnostic emotion vector.}
  \label{fig:training_process}
  \vspace{-10pt} %
\end{figure}

\section{Related work}
\subsection{Task Arithmetic}
Task arithmetic~\cite{task_arithmetic} is one of the model editing techniques that modifies a model's behavior by directly manipulating model parameters via a task vector, without additional training.
The task vector is defined as \mmedit{the difference between the model parameters before and after fine-tuning.}
These vectors are scaled and combined via arithmetic operations to modify the model's capabilities and improve performance on various tasks.
Previous studies have shown that task arithmetic can improve the accuracy of classification and recognition tasks by applying the task vectors~\cite{model_soup, merge_asr}. 
Moreover, it enables the acquisition of multitasking skills by combining task vectors from different tasks~\cite{task_arithmetic, ties_merge} and enables the generation of the output with intermediate attributes in both image and speech synthesis~\cite{gan_cocktail, merge_attribute}.
Unlike traditional transfer learning methods that require additional training, task arithmetic directly modifies existing model parameters without training. This approach reduces computational costs, simplifies the process of adapting models to new tasks, and enhances their performance.

\subsection{Emotion Arithmetic}
\label{subsection:emotion_arithmetic}
Emotion arithmetic was proposed by Kalyan et al.~\cite{emotion_arithmetic}, which applied the task arithmetic method to emotion TTS models for controling emotion intensity.
They began by creating a pre-trained TTS model trained on a neutral style single-speaker speech dataset.
Subsequently, they fine-tuned this pre-trained model using the emotion speech data of the same speaker.
Then, the task vector, referred to as the ``emotion vector'', was obtained by subtracting the pre-trained model parameters from the fine-tuned ones.
Finally, the emotion vector was scaled and added to the arbitrary single-speaker pre-trained TTS model to control emotion intensity, as follows:
\begin{align}
\label{equation:model_merge}
\boldsymbol{\tau}^{\mathtt{spkA}}_\mathtt{emo} &= \boldsymbol{\theta}^{\mathtt{spkA}}_{\mathtt{emo}} - \boldsymbol{\theta}^{\mathtt{spkA}}_{\mathtt{pre}}, \\
\boldsymbol{\theta}^{\mathtt{target\_spk}}_{\mathtt{new}} &= \boldsymbol{\theta}^{\mathtt{target\_spk}}_{\mathtt{pre}} + \alpha \cdot \boldsymbol{\tau}^{\mathtt{spkA}}_\mathtt{emo},
\end{align}
where $\boldsymbol{\tau}^{\mathtt{spkA}}_\mathtt{emo}$ denotes the emotion vector of speaker A, and 
$\boldsymbol{\theta}^{\mathtt{spkA}}_{\mathtt{pre}}$ represents the model parameters of the pre-trained TTS model trained on the neutral speech data of speaker A. 
$\boldsymbol{\theta}^{\mathtt{spkA}}_{\mathtt{emo}}$ is the fine-tuned model parameters of the emotion TTS model trained on the emotional speech data of speaker A.
$\boldsymbol{\theta}^{\mathtt{target\_spk}}_{\mathtt{pre}}$ is the model parameters of the pre-trained TTS model of the target speaker.
The same (target speaker = speaker A) or a different speaker model (e.g. speaker B, or speaker C) can be used as the target speaker pre-trained model.
$\alpha$ denotes the scaling factor of the emotion vector for controlling the desired emotion intensity.
$\boldsymbol{\theta}^{\mathtt{target\_spk}}_{\mathtt{new}}$ is the resultant model parameters.

Although this method achieved promising results in the same-speaker settings, their experimental results showed that the generated speech quality and speaker consistency are degraded in the cross-speaker setting.
In this study, we aim to address this degradation issue in the cross-speaker setting.

\begin{figure}[t]
   \centering
    \begin{minipage}[b]{1\linewidth}
      \centering
      \includegraphics[width=\hsize]{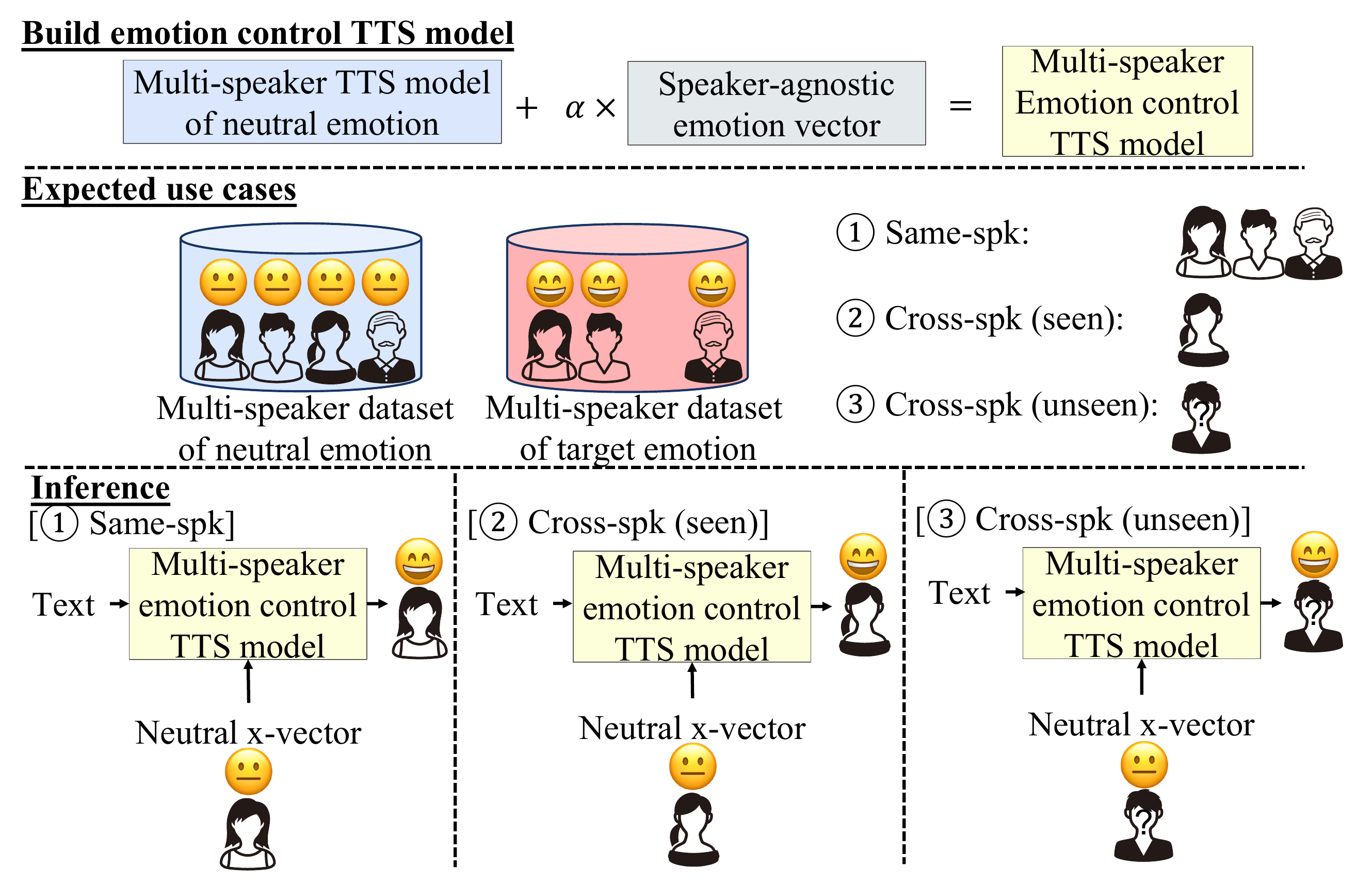}
    \end{minipage}
    \caption{Inference process in three expected use cases.}
    \vspace{-10pt} %
    \label{fig:inference_process}
  \end{figure}

\section{Proposed method}
\label{section:proposed_method}

To address the limitations of a single-speaker emotion vector in cross-speaker emotion intensity control, we propose a speaker-agnostic emotion vector. 
This approach aims to improve both speech quality and speaker consistency during emotion intensity control in the cross-speaker case.

Figure~\ref{fig:training_process} illustrates the process of creating
the proposed speaker-agnostic emotion vector.
Unlike emotion arithmetic, which used a single-speaker model, we first trained an x-vector-based multi-speaker TTS model as the pre-trained model
using neutral speech data from multiple speakers.
\mmedit{During pre-training, we assume the training dataset included two types of speakers: (1) speakers with both neutral and emotional speech data and (2) speakers with only neutral speech data.}
Subsequently, we fine-tuned the pre-trained model on multi-speaker emotion speech data.
During the fine-tuning process, the model was conditioned on each speaker's neutral x-vector, consistent with the pre-training process.
The speaker-agnostic emotion vector $\boldsymbol{\tau}_\mathtt{emo}^{\mathtt{multi}}$ was then obtained by subtracting the pre-trained neutral TTS model parameters $\boldsymbol{\theta}^{\mathtt{multi}}_\mathtt{{pre}}$ from the fine-tuned emotion TTS model parameters $\boldsymbol{\theta}^{\mathtt{multi}}_{\mathtt{emo}}$.
We expect this speaker-agnostic emotion vector $\boldsymbol{\tau}_\mathtt{emo}^{\mathtt{multi}}$ to capture speaker-independent emotion representations. By scaling the speaker-agnostic emotion vector $\boldsymbol{\tau}_\mathtt{emo}^{\mathtt{multi}}$ and adding it to the pre-trained model parameters $\boldsymbol{\theta}_{\mathtt{pre}} ^{\mathtt{multi}}$, we obtain new resultant model parameters $\boldsymbol{\theta}_{\mathtt{new}}^{\mathtt{multi}}$, as follows:
\begin{align}
\label{equation:model_merge}
\boldsymbol{\tau}_\mathtt{emo}^{\mathtt{multi}} &= \boldsymbol{\theta}^{\mathtt{multi}}_{\mathtt{emo}} - \boldsymbol{\theta}^{\mathtt{multi}}_\mathtt{{pre}}, \\
\boldsymbol{\theta}_{\mathtt{new}}^{\mathtt{multi}} &= \boldsymbol{\theta}_{\mathtt{pre}} ^{\mathtt{multi}}+ \alpha  \cdot \boldsymbol{\tau}_\mathtt{emo}^{\mathtt{multi}},
\end{align}
where $\alpha$ denotes the scaling factor for the speaker-agnostic emotion vector.
This resulting model enables cross-speaker emotion intensity control.

\subsection{Speaker conditioning for multi-speaker TTS model}
\label{sec:model_merging_method}

\mmedit{In practical usage, obtaining specific emotional speech data from a target speaker is not always easy.
Therefore, our proposed method aims to generate the emotional speech of the target speakers by using only their neutral speech.}

\mmedit{To achieve this, we fine-tuned the model to generate emotional speech of the target speaker using their neutral speech conditioning.
We then expect the model to acquire a transformation from neutral speech conditioning into their emotional speech representation.
To ensure consistency, 
we used the same speaker conditioning derived from their neutral speech for both the pre-training and fine-tuning processes.
Specifically, the speaker x-vector for each speaker is used as speaker conditioning.
We then created the speaker x-vectors by averaging the x-vectors obtained from each speaker's neutral speech.}
During the inference process, the model was conditioned on the target speaker's x-vector, extracted from their neutral reference speech, as shown in Figure~\ref{fig:inference_process}.

\subsection{Expected use cases}
\label{subsection:expected_cases}

Figure~\ref{fig:inference_process} illustrates the inference process in the three expected use cases.
In this study, we assume three use cases for emotion intensity control based on the availability of emotion speech data for the target speaker, as follows:

\noindent\textbf{Same-spk}: 
In this ``same-speaker'' case, we synthesize emotional speech of the ``seen'' target speaker with the desired emotion intensity.
Both the neutral and emotional speech data from the target speaker are available during training.

\noindent\textbf{Cross-spk (seen)}:
In this ``seen cross-speaker'' case, only the neutral speech data of the seen target speaker is available during training, without using their emotional speech data.
This case aims to generate emotional speech of the ``seen'' speaker with the desired emotion intensity.

\noindent\textbf{Cross-spk (unseen)}:
In this ``unseen cross-speaker'' case, neither neutral nor emotional speech data from the target speaker is used during training (i.e., a zero-shot speaker situation).
This case aims to generate emotional speech of an ``unseen'' speaker with the desired emotion intensity.

\section{Experiments}
\subsection{Experimental settings}

\subsubsection{Datasets}
\label{subsubsection:datasets}

In this study, we used two datasets: emotional speech dataset (ESD) and VCTK.

\noindent\textbf{ESD}: ESD~\cite{zhou2021emotional,zhou2021seen} dataset contains approximately 29 h of English speech from 10 native speakers with five emotion styles, including neutral style.
In the proposed method, the neutral speech data were used for pre-training, while three emotion-style speech data (Angry, Sad, and Happy)  were used as the emotion speech data for fine-tuning.

\noindent\textbf{VCTK}: The VCTK~\cite{vctk} dataset consists of approximately 44 h of neutral-style English speech from 109 speakers. We used 100 speakers as seen speakers for pre-training and selected 4 out of 9 rest speakers as unseen speakers for evaluation.

A multi-speaker neutral TTS pre-trained model was created using the seen speaker subset from VCTK and the neutral speech subset from ESD. Subsequently, we fine-tuned this model using the emotion speech from all speakers in ESD, starting with the pre-trained model initialization.
For both datasets, we created training/validation/test splits of approximately 90\%/5\%/5\% of the original dataset, respectively.

\subsubsection{Model architecture}

The proposed method adopted the implementation of Conformer-FastSpeech2 (CFS2)~\cite{cfs2, conformer, fastspeech2} from ESPnet~\cite{espnet}.
We conditioned the text encoder output on the speaker x-vector~\cite{xvector}, which was extracted using the pre-trained speaker encoder from SpeechBrain~\cite{speechbrain}.
For waveform generation, we used a pre-trained HiFi-GAN~\cite{hifigan} vocoder from the ParallelWaveGAN repository\footnote{\url{https://github.com/kan-bayashi/ParallelWaveGAN}} trained on the VCTK dataset.

\subsubsection{Baseline method}

We used the single-speaker emotion vector from the original emotion arithmetic method as the baseline.
For a fair comparison, we adopted the same CFS2 architecture used in the baseline method.
We denote the single-speaker emotion vector approach as the ``baseline.''

\subsubsection{Target speaker settings for evaluation}
We evaluated the performance of the proposed method on the emotion intensity control task across the three expected use cases described in Section~\ref{subsection:expected_cases}.

For the same-speaker case (\textbf{Same-spk}), we selected 4 speakers (2 male and 2 female) from the ESD (Speaker IDs: 0012, 0014, 0015, and 0016) for evaluation.

For the seen cross-speaker case (\textbf{Cross-spk (seen)}), we selected 4 speakers (2 male and 2 female) from the seen speaker subset of VCTK (Speaker IDs: p226, p228, p232, and p234) as ``seen'' target speakers.

For the unseen cross-speaker case (\textbf{Cross-spk (unseen)}), we selected 4 speakers (2 male and 2 female) from the unseen speaker subset of the VCTK dataset (Speaker IDs: p361, p362, p363, and p374) as ``unseen'' target speakers.

\subsection{Speech quality evaluation}
\label{sec:mos_wer_test}

To assess the speech quality across the three target speaker cases described in Section~\ref{subsection:expected_cases}, we conducted a mean opinion score (MOS) evaluation to investigate the naturalness of the synthesized emotional speech.

In the MOS evaluation, participants rated the naturalness of the given speech samples on a five-point scale. 
For each sample, we randomly selected a target speaker and generated their emotional speech samples with an emotion intensity scaling factor of $\alpha=0.9$. 
We collected 200 responses per emotion for each method.
Participants evaluated five different samples, each based on 10 randomly selected sentences from the VCTK test subset. All participants were recruited through the crowdsourcing platform Amazon Mechanical Turk (AMT)\footnote{\url{https://www.mturk.com/}}. 
The MOS scores were calculated by averaging the ratings across speakers for each method.

\subsection{Speaker consistency evaluation}
\label{sec:mos_wer_test}

To evaluate how well speaker identity is preserved in the synthesized emotional speech during emotion intensity control, we calculated the speaker encoder cosine similarity (SECS)~\cite{secs} between the x-vectors of the emotional ($\alpha=0.9$) and neutral synthetic speech from the corresponding target speaker. Each speech sample was synthesized using 10 sentences from the VCTK test set.
The SECS scores were calculated by averaging the similarity scores across speakers for each method.

\subsection{Cross-speaker emotion intensity controllability}
\label{sec:secs_evaluation}

To assess the cross-speaker emotion intensity controllability, we conducted a rearrangement experiment to evaluate how accurately listeners can distinguish synthetic emotional speech with varying intensities in both seen and unseen target speaker cases.
The synthesized emotion speech included three different intensity levels, $\alpha=0.1$ (weak), $0.5$ (medium), and $0.9$ (strong).
Participants were recruited through AMT and asked to sort randomly shuffled three speech samples of the same sentence, each with a different emotion intensity, according to their perceived emotional intensity. 
We collected 200 responses per method for each emotion.

\section{Experimental results}
Table~\ref{table:mos_evaluation} shows the MOS and SECS scores for each target emotion across the three different cases described in Section~\ref{subsection:expected_cases}.
Figure~\ref{fig:confusion_matrix} shows the confusion matrices for each target emotion with different intensity levels. The diagonal values in these matrices represent how accurately the intended emotion intensity of the synthesized speech was perceived.

\subsection{Speech quality evaluation}
In the same-speaker case, the proposed method achieved better naturalness compared to the baseline across all emotions. While the baseline method only used single-speaker emotion speech data for training,
our proposed method used a multi-speaker emotion speech dataset, leading to the improvement of the synthesized emotional speech quality for each target speaker.

In the seen cross-speaker case, prior research has reported that while the baseline method could generate comprehensive speech, it often failed to preserve speaker consistency. 
However, in our reproduced implementation, the baseline method was unable to generate comprehensible speech during this experiment.
In contrast, our proposed method could generate emotion speech with comparable speech naturalness to the same-speaker case, while the emotional speech data of the target speaker were not included in the training data.

In the unseen cross-speaker case, the proposed method successfully achieved comparable synthesized speech quality to that of the seen cross-speaker case. 
This indicates that our proposed method has the zero-shot capability to synthesize emotional speech with high quality.

\begin{table}[t!]
    \begin{center}
    \caption{The MOS and SECS score with 95\% confidence interval (CI) for each three emotion style.}
    \scalebox{0.62}{
        \begin{tabular}{ c|c|cc|c|c } 
            \toprule
            \multicolumn{6}{c}{MOS ($\uparrow$)} \\
            \toprule
             \multirow{2}{*}{Emotion} &\multirow{2}{*}{GT} &\multicolumn{2}{c|}{Same-spk} & \multicolumn{1}{c|}{Cross-spk (seen)} &  \multicolumn{1}{c}{Cross-spk (unseen)}\\ 
             & & baseline & proposed & proposed & proposed \\
            \midrule
            Angry &  3.85 $\pm$ 0.16  &  2.32 $\pm$ 0.15  &  \textbf{3.78 $\pm$ 0.14} & 3.69 $\pm$ 0.15 & 3.73 $\pm$ 0.16\\
            Sad &  3.93 $\pm$ 0.14 & 2.27 $\pm$ 0.14 & \textbf{3.54 $\pm$ 0.13} & 3.88 $\pm$ 0.12 & 3.85 $\pm$ 0.14\\
            Happy &  4.01 $\pm$ 0.15  &  2.44 $\pm$ 0.16  & \textbf{3.83 $\pm$ 0.13} & 3.92 $\pm$ 0.13  & 3.81 $\pm$ 0.14\\
            \toprule
            \multicolumn{6}{c}{SECS ($\uparrow$)} \\
            \toprule
             \multirow{2}{*}{Emotion} &\multirow{2}{*}{GT} &\multicolumn{2}{c|}{Same-spk} & \multicolumn{1}{c|}{Cross-spk (seen)} &  \multicolumn{1}{c}{Cross-spk (unseen)}\\ 
             & & baseline & proposed & proposed & proposed \\
            \midrule
            Angry &  0.80 $\pm$ 0.05 & \textbf{0.88 $\pm$ 0.05} & \textbf{0.88 $\pm$ 0.03} & 0.78 $\pm$ 0.06 & 0.80 $\pm$ 0.06 \\
            Sad & 0.78 $\pm$ 0.07  & 0.81 $\pm$ 0.10 & \textbf{0.88 $\pm$ 0.05} & 0.80 $\pm$ 0.05 & 0.84 $\pm$ 0.05 \\
            Happy & 0.78 $\pm$ 0.06  & \textbf{0.87 $\pm$ 0.06} & 0.85 $\pm$ 0.04 & 0.78 $\pm$ 0.06 &  0.79 $\pm$ 0.07 \\
            \bottomrule
        \end{tabular}
    }
    \label{table:mos_evaluation}
    \end{center}
          \vspace{-20pt} %
\end{table}

\subsection{Speaker consistency evaluation}

\mmedit{In the same-speaker case, the proposed method consistently achieved SECS scores above 0.85 for all emotions (Angry, Sad, and Happy), indicating that our method achieved comparable or slightly better speaker consistency than that of the baseline method.}

\mmedit{In the cross-speaker case, while the prior research reported that the baseline method lost speaker consistency, the proposed method achieved SECS scores ranging from 0.78 to 0.84, which is comparable or slightly better scores than the ground truth range of 0.78 to 0.80. 
Even in the unseen cross-speaker case, our proposed method achieved comparable speaker consistency to that of the ground truth.}

\mmedit{While the SECS scores between emotional and neutral ground truth speech range from 0.78 to 0.80, the proposed method consistently achieved SECS scores above 0.78 across all cases.} 
\mmedit{These results suggest that the proposed method effectively preserves speaker consistency between neutral speech and emotional speech even for unseen target speakers.}

\subsection{Cross-speaker emotion intensity controllability}
\begin{figure}[t]
  \begin{minipage}[b]{1.0\linewidth}
    \centering
    \includegraphics[width=\hsize]{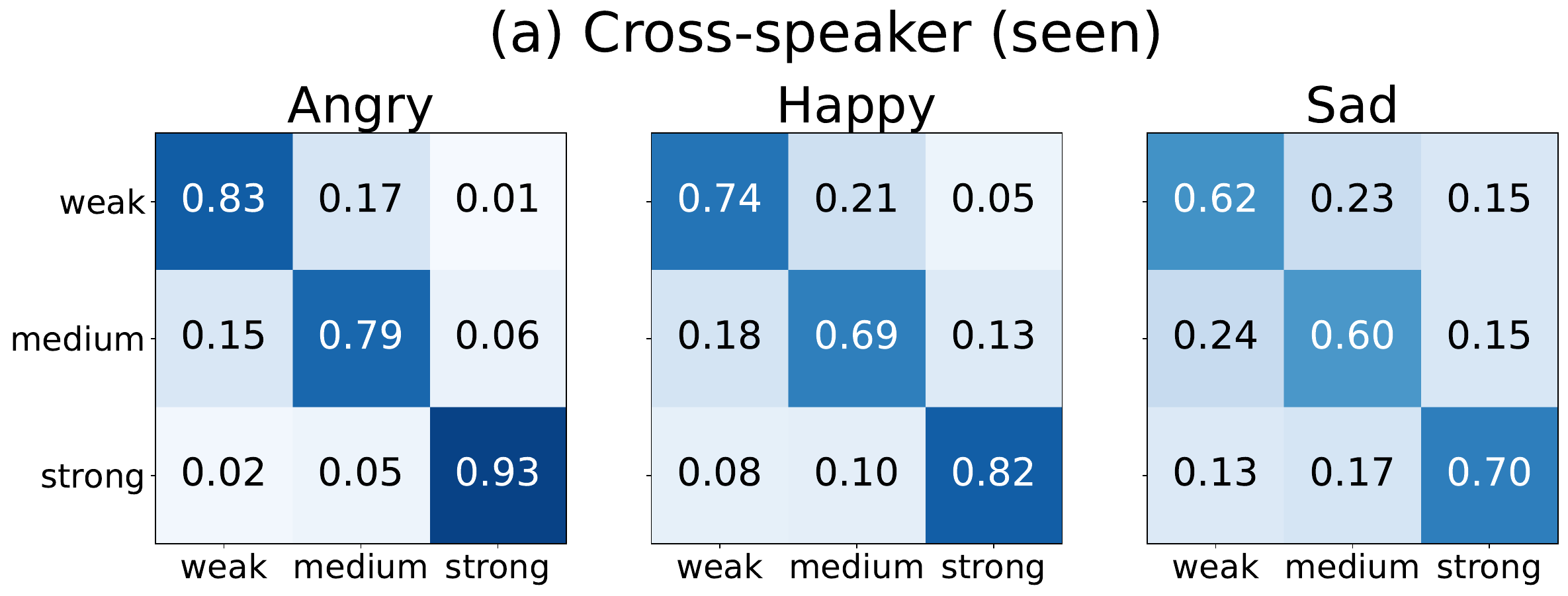}
  \end{minipage}
    \begin{minipage}[b]{1.0\linewidth}
    \centering
\includegraphics[width=\hsize]{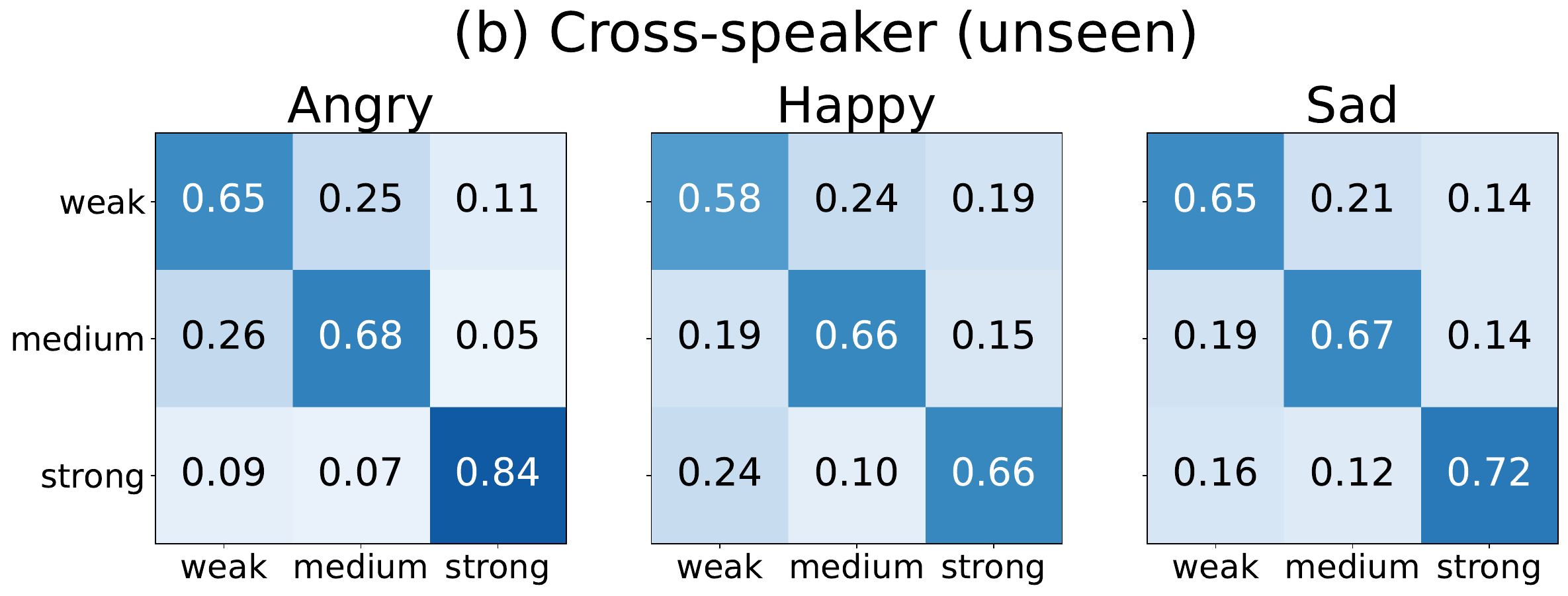}
  \end{minipage}
  \caption{Confusion matrix of synthesized emotion speech from the proposed method in both (a) Cross-speaker (seen) and (b) Cross-speaker (unseen) cases. 
  The vertical and horizontal axis of each figure represent perceived and ground-truth emotion strength, respectively.}
  \label{fig:confusion_matrix}
      \vspace{-10pt} %
\end{figure}

Figure~\ref{fig:confusion_matrix} shows that the proposed method achieved a mean perceived emotion intensity accuracy of 0.74 in the seen cross-speaker case and 0.67 in the unseen cross-speaker case. This indicates that participants were able to distinguish varying emotion intensities.
These results show that the speaker-agnostic emotion vector has the controllability of emotion intensities at a perceivable level.

\section{Conclusions}
\label{sec:conclusion}

In this study, we proposed a speaker-agnostic emotion vector for cross-speaker emotion intensity control. 
The proposed speaker-agnostic emotion vector was designed to capture the shared emotional expressions across multiple speakers.
We conducted both subjective and objective evaluations on the emotion intensity control task from various perspectives, such as naturalness, speaker consistency, and emotion intensity controllability.
Our experimental results showed that the proposed method achieves perceptible emotion intensity controllability in the cross-speaker case while maintaining its speech quality and speaker consistency, even for the unseen speaker.

In future work, we plan to explore the capability of the proposed method to handle mixed emotion cases even if the target emotion speech data is not available.

\bibliographystyle{IEEEtran}
\bibliography{mybib}

\end{document}